%Paper: cond-mat/9509111
%From: janco@stat.th.u-psud.fr (jancovici)
%Date: Tue, 19 Sep 1995 12:17:30 +0100

\magnification 1200  \baselineskip=18pt
 \hfuzz=5pt \null\vskip 2cm
\rightline{{\it To the memory of Claude Itzykson}}
\bigskip
\centerline{\bf TWO-DIMENSIONAL ONE-COMPONENT PLASMA}
\centerline{\bf  IN A QUADRUPOLAR FIELD}
\bigskip
\centerline{P.J. FORRESTER\footnote{*}{E-mail : matpjf@maths.mu.oz.au}
 and B. JANCOVICI\footnote{**}{Permanent address : Laboratoire de
Physique Th\'eorique et Hautes Energies, Universit\'e de Paris-Sud,
91405 Orsay, France (Laboratoire associ\'e au Centre National de la
Recherche Scientifique). E-mail : janco@stat.th.u-psud.fr}}
\smallskip
\centerline{Department of Mathematics, University of Melbourne,
Parkville, Victoria 3052, Australia}
\vskip 1cm

\noindent The classical two-dimensional one-component plasma is an
exactly solvable model, at some special temperature, even when the
one-body potential acting on the particles has a quadrupolar term.
As a supplement to a recent work of Di Francesco, Gaudin, Itzykson,
and Lesage [{\it Int. J. Mod. Phys.} {\bf A9}, 4257 (1994)] about an
$N$-particle system ($N$ large but finite), a macroscopic argument is
given for confirming that the particles form an elliptical blob, the
analogy between the classical plasma and a quantum $N$-fermion system
in a magnetic field is used for the microscopic approach, and a
microscopic calculation of the surface charge-surface charge
correlation function is performed; an expected universal form is
shown to be realized by this correlation function.
\vskip 1cm

\noindent{\obeylines LPTHE Orsay 95/36
\noindent May 1995}
\vfill\eject
\noindent {\bf 1. Introduction}
\bigskip
In a recent paper published in this journal$^1$, Di Francesco,
Gaudin, Itzykson, and Lesage have studied a  classical two-dimensional
one-component plasma submitted to a quadrupolar electric field. The
present paper is a supplement to the work of these authors.

The model under consideration consists of $N$ classical particles of
charge $e$ immersed in a uniform neutralizing background. The
particles interact through the two-dimensional Coulomb potential
$-(e^2/2\pi) \ln |z_i - z_j|$, where the position of the {\it i}-th
particle is defined by the complex number $z_i$; there are also
background particle and background-background interactions. A
dimensionless coupling constant is $\beta = e^2/2\pi k_BT$, where $T$
is the temperature and $k_B$ is Boltzmann's constant. For the special
value $\beta = 2$, the model is exactly solvable in a variety of
geometries$^{2-5}$. A simple case is when the system is a disk; then
the background generates a one-body potential of circular symmetry.
However, Di Francesco {\it et al} have considered the more general
case when the one-body potential $V(z, \bar z)$ also has a
quadrupolar component ; up to some
additive constant,
$$
{V(z, \bar z)\over k_BT} = z \bar z - {1\over 2} (\tanh \mu) (z^2 +
\bar z^2) . \eqno (1.1)
$$
With this choice, the model can still be solved for $\beta = 2$. Di
Francesco {\it et al} have studied the density of the $N$-particle
system in the large-$N$ limit, and they found that the particles form
a blob of constant density $1/\pi$, having the shape of an ellipse
with semi-axes $a = \sqrt{N} e^\mu$ and $b = \sqrt{N} e^{-\mu}$.

In the present paper, we revisit the $N$-particle system in a
quadrupolar field. In Section 2, we retrieve the elliptic shape of
the particle blob by a simple macroscopic argument. In Section 3, we
revisit the microscopic approach by using the analogy with a system
of independent fermions in a magnetic field. In Section 4, we study
the correlations near the boundary of the blob.
\bigskip
\noindent{\bf 2. Macroscopic equilibrium}
\bigskip
The one-body potential (1.1) can be thought of as generated by a
uniform background with the shape of a large ellipse. Indeed, if the
background charge density is $-e/\pi$ and the ellipse has semi-axes
$A$ and $B$ such that $B/A = e^{-2\mu}$, a simple calculation$^{6}$
shows that, at a point $z$ inside the ellipse, this background
generates an electric potential $V/e$ given by (1.1), up to some
additive constant (here $k_BT = e^2/4\pi)$.

Let us now put  $N$ particles in, assuming $N << AB$ (this condition
will ensure that the particles stay well inside the background). The
particles will form a blob, and macroscopic electrostatics requires
that the total electrostatic potential be a constant inside that
conducting blob. Therefore, the density and shape of the blob should
be such that it generates an electric potential $-V/e$, up to an
additive constant. The same calculation as above says that the blob
which achieves that has the shape of an ellipse, with semi-axes $a$
and $b$ such that $b/a = e^{-2\mu}$, and a constant particle density
$N/\pi ab = 1/\pi$. Thus, one retrieves the result of Di Francesco
{\it et al} : $a = \sqrt{N} e^\mu$, $b = \sqrt{N} e^{-\mu}$.
\bigskip
\noindent {\bf 3. Magnetic picture}
\bigskip
The Boltzmann factor of the one-component plasma is, up to a
multiplicative constant,
$$
\exp\Big \{ \sum^N_{j=1} \big [-z_j \bar z_j + {1\over 2} (\tanh
\mu)(z^2_j + \bar z_j^2)\big ]\Big \} |\Delta|^2 \eqno (3.1)
$$
where
$$
\Delta = \det \left |\matrix{1 & z_1 & z^2_1 \ldots &
z^{N-1}_1\cr 1 & z_2 & z^2_2 \ldots & z^{N-1}_2 \cr  .. & ..
& ....... & ...\cr 1 & z_N & z^2_N \ldots & z^{N-1}_N \cr}\right | =
\prod_{i > j}(z_i - z_j) .
$$
Di Francesco {\it et al} have expressed this Boltzmann factor in
terms of {\it orthogonal} functions. We now rephrase their
work, in a way that we believe to be more transparent, using the
magnetic language. Indeed, in the special case of circular symmetry
$(\mu = 0)$, (3.1) was introduced as a squared wavefunction of a
quantum system : $N$ independent fermions in a plane, interacting
with a uniform magnetic field normal to the plane. Let us first
review this picture$^7$ for $\mu = 0$.

The quantum Hamiltonian for a particle of mass $1$ and charge $1$ in
a magnetic field is $H = (1/2)({\bf p} - {\bf A})^2$, where ${\bf p}$
is the canonical momentum and ${\bf A}$ the vector potential. For
describing a uniform magnetic field normal to the plane (the value
of which is chosen  as $2$), a possible gauge is $A_x = -y, A_y = x$
($x$ and $y$ are Cartesian coordinates in the plane; $z = x + iy$).
In terms of annihilation and creation operators for ``left
quanta''$^7$
$$
\eqalign{a &= {1\over 2} (x + iy + ip_x - p_y) = {1\over 2} (x + iy +
\partial_x + i\partial_y) = {1\over 2} z + \partial_{\bar z} \qquad
, \cr
 a^+ & = {1\over 2} (x - iy - ip_x - p_y) = {1\over 2} (x - iy -
\partial_x + i\partial_y) = {1\over 2} \bar z - \partial_z
\qquad , \cr}
$$
the Hamiltonian is\footnote{$^a$}{It is convenient here to use sign
conventions different from the ones in ref. 7.}
$$
H = 2 a^+ a + 1   .
$$
For obtaining a complete set of commuting  observables, we must also
introduce annihilation and creation operators for ``right quanta''
$$
\eqalign{b &= {1\over 2} (x - iy + ip_x + p_y) = {1\over 2} (x - iy +
\partial_x - i\partial_y) = {1\over 2} \bar z + \partial_ z
\qquad , \cr
 b^+ & = {1\over 2} (x + iy - ip_x + p_y) = {1\over 2} (x + iy -
\partial_x - i\partial_y) = {1\over 2} z - \partial_{\bar z} \qquad
. \cr}
$$
The only  non-vanishing commutators betwen these operators are
$$
[a, a^+] = 1 \qquad , \qquad [b, b^+] = 1 \ ,
$$
and $a^+a$ and $b^+b$ do form a complete set of commuting
observables.The wavefunctions $\psi$ associated to the lowest Landau
level obey $a \psi = 0$. They can be labeled by the eigenvalue
$\ell$ of the angular momentum $L = b^+b - a^+a$ which, for the
lowest Landau level, reduces to $b^+b$. Starting from $\psi_0$
defined by
$$
a \psi_0 = 0 \qquad \qquad, b \psi_0 = 0 \ ,
$$
the normalized solution of which is
$$
\psi_0 = {1\over \sqrt{\pi}} e^{-{1\over 2} z \bar z} \qquad ,
$$
we can obtain the other states of the lowest Landau level by
applying the creation operator $b^+$, with the result
$$
\psi_\ell = {1 \over \sqrt{\ell !}} (b^+)^\ell \psi_0 = {1\over
\sqrt{\pi \ell !}} z^\ell e^{-{1\over 2} z\bar z} \qquad . \eqno (3.2)
$$
For $\mu = 0$, the Boltzmann factor (3.1) is, up to a multiplicative
factor, the squared Slater determinant built with the
orthogonal wavefunctions $\psi_\ell, 0 \leq \ell \leq N - 1$.

In the general case $\mu \not = 0$, the Boltzmann factor (3.1) can
again be expressed as a squared Slater determinant, built with
orthogonal wave functions, for the same magnetic problem as above. Now
the normalized wavefunction $\psi_0$ has to be replaced by
$$
\tilde \psi_0 = {1\over \sqrt{\pi \cosh \mu}} e^{-{1\over 2} z \bar z
+ {1\over 2} (\tanh \mu) z^2} \quad . \eqno (3.3)
$$
Since $\tilde \psi_0$ can be expanded in powers of $\tanh \mu$ as a
linear combination of the previous functions (3.2), it is another
lowest Landau level wavefunction, which obeys
$$
a  \tilde \psi_0 = 0 \quad .
$$
The operators $a$ and $a^+$ are unchanged. However, for generating
the other mutually orthogonal functions which enter the Slater
determinant, we must replace $b$ and $b^+$ by $\tilde b$ and $\tilde
b^+$ such that
$$
\tilde b \tilde \psi_0 = 0 \eqno (3.4)
$$
and such that the commutation relations be preserved. This is easily
achieved by a Bogolyubov transformation, i.e. writing for $\tilde b$
a linear combination of $b$ and $b^+$, the coefficients of which are
fixed (up to an irrelevant phase) by the conditions (3.4) and
$[\tilde b, \tilde b^+] = 1$. One finds
$$
\tilde b^+ = (\cosh \mu) b^+ - (\sinh \mu) b = (\cosh \mu) ({1\over
2} z - \partial_{\bar z}) -(\sinh \mu) ({1\over 2} \bar z +
\partial_z) \eqno (3.5)
$$
and the orthogonal wave functions are
$$
\tilde \psi_n = {1\over \sqrt{n !}} (\tilde b^+)^n \tilde \psi_0 .
\eqno (3.6)
$$
Using (3.3) and (3.5) in (3.6), one readily sees that the $\tilde
\psi_n$ are of the form
$$
\tilde \psi_n = P_n(z) e^{- {1\over 2} z\bar z + {1\over 2} (\tanh
\mu)z^2}
$$
where $P_n(z)$ is a polynomial of degree $n$ defined by the
recurrence relation
$$
P_{n+1}(z) = {1\over \sqrt{n+1}} \Big [ {1\over \cosh \mu} z P_n(z)
- (\sinh \mu) {dP_n(z)\over dz}\Big ] \eqno (3.7)
$$
 and $P_0(z) = (\pi \cosh \mu)^{-1/2}$. A simple change of variable
relates (3.7) to the recurrence relation$^8$ obeyed by the Hermite
polynomials\footnote{$^b$}{We define the Hermite polynomials as in
ref.8, with a weight function $\exp(-x^2)$, while Di Francesco {\it
et al} used $\exp(-x^2/2)$.} $H_n(x)$,
$$
H_{n+1}(x) = 2x H_n(x) - {dH_n(x)\over dx}
$$
and $H_0(x) = 1$, and one obtains
$$\eqalign{
P_n(z) &= {1\over \sqrt{n !}} ({1\over 2} \tanh \mu)^{n/2} (\pi \cosh
\mu)^{-1/2}  H_n({z\over \sqrt{sh2\mu}})\cr
&= (\pi n!)^{-1/2} (\cosh \mu)^{-n-{1\over 2}} z^n + \ldots  \
.\cr}
$$
 Therefore, by making linear combinations of the columns in the
determinant $\Delta$ of (3.1), one can replace each $z^n$ by $(\pi
n!)^{1/2} (\cosh \mu)^{n+{1\over 2}} P_n(z)$, and (3.1) becomes a
constant times the squared modulus of a Slater determinant built with
orthonormal wavefunctions. This is another derivation of the result
obtained by Di Francesco {\it et al}.
\bigskip
\noindent {\bf 4. Surface correlations}
\bigskip
Thus, the density correlation functions associated to the Boltzmann
factor (3.1) are the same ones as for $N$ fermions occupying the
orthonormal orbitals $\tilde \psi_n (n = 0, 1, \ldots N-1)$.  All the
$n$-point densities can be expressed in terms of the projector
$$
\eqalignno{
K_N({\bf r}, {\bf r'}) &= \sum_{n=0}^{N-1} \tilde \psi_n ({\bf r})
\bar{\tilde \psi}_n({\bf r'})\cr
&= {1\over \pi} a({\bf r}) \bar a({\bf r'}) \sum^{N-1}_{n=0} {(\tanh
\mu)^n\over (\cosh \mu) n !} 2^{-n} H_n ({z\over \sqrt{\sinh 2\mu}})
H_n ({\bar z'\over \sqrt{\sinh 2 \mu}}) &(4.1)\cr}
$$
where
$$
a({\bf r}) = e^{- {1\over 2} z \bar z + {1\over 2} (\tanh \mu) z^2}
\qquad .
$$
In particular, the density at ${\bf r}$ is $K_N({\bf r}, {\bf r})$
and the two-point truncated density (correlation function) is
$$
\rho_N^T({\bf r}, {\bf r'}) = - | K_N({\bf r}, {\bf r'})|^2 \qquad .
\eqno (4.2)
$$

The particles form an elliptical blob. The correlations are expected
to be long-ranged only near the boundary$^{9,10}$ and we can define a
surface density-surface density correlation by integrating the
long-range part of $\rho^T_N({\bf r}, {\bf r'})$ with respect to the
components of ${\bf r}$ and ${\bf r'}$ normal to the boundary of the
blob. Our aim is to evaluate that correlation function for macroscopic
distances.

It is appropriate to work with elliptic coordinates $(\xi, \eta)$
related to the Cartesian coordinates $(x, y)$ by
$$
x + iy = [(2N - 1) \sinh 2 \mu]^{1/2} \cosh (\xi + i \eta) \qquad .
\eqno (4.3)
$$
This choice ensures that the boundary ellipse (see Section 2) is one
of the coordinate lines, $\xi = \xi_b$, where $e^{-2\xi_b} = \tanh
\mu$, up to a negligible correction of order $1/N$ (using $2N-1$
rather than $2N$ in the definition (4.3) will turn out to be
convenient in the following). The rescaling factor of the conformal
transformation (4.3) is
$$
h(\xi, \eta) = | {d(x + iy)\over d(\xi + i \eta)}| = [(2 N-1) \sinh 2
\mu]^{1/2} | \sinh (\xi + i \eta)|\qquad , \eqno (4.4)
$$
and the surface charge-surface charge correlation is
$$
\langle \sigma(\eta) \sigma(\eta')\rangle^T = e^2 \int^\infty_0 d\xi
h(\xi, \eta) \int^\infty_0 d\xi' h(\xi', \eta') \rho_N^{T(b)} ({\bf
r}, {\bf r'})\qquad, \eqno (4.5)
$$
where $\rho_N^{T(b)}({\bf r}, {\bf r'})$ is the asymptotic form of
$\rho^T_N({\bf r}, {\bf r'})$ when $|{\bf r} - {\bf r'}|$ is a
macroscopic distance (it is non-vanishing only if ${\bf r}$ and
${\bf r'}$ are near the boundary).

The surface charge-surface charge correlation is particularly
interesting as its asymptotic form is expected to depend on the shape
of the plasma, but to be otherwise universal (i.e. independent of the
microscopic detail).$^{10}$ For an ellipse-shaped plasma this form
is$^{11}$
$$
{1\over 2\pi k_BT} \langle \sigma(\eta) \sigma (\eta') \rangle^T \sim
- {1\over 2 \pi^2 |e^{i\eta}- e^{i\eta'}|^2 h(\xi_b, \eta) h(\xi_b,
\eta')} \qquad .\eqno (4.6)
$$
The exact result (4.2) therefore provides an opportunity to test
(4.6).

To evaluate (4.5) and thus test (4.6), our major task is to compute
$\rho^{T(b)}_N$. For our overall strategy in this calculation, we
take guidance from the calculation of Choquard {\it et al},$^{12}$ who
calculated $\rho^{T(b)}_N ({\bf r}, {\bf r'})$ in disk geometry. Also,
we make use of the known$^{13}$ large-$n$ asymptotic expansion for
the Hermite polynomial $H_n(x)$  :
$$
\eqalignno{
H_n(x) \sim &{2^{n/2-1} e^{[n+x^2 - x(x^2-2n-1)^{1/2}]/2} n!\over
\pi^{1/2} n^{(n+1)/2} ({x^2\over 2n+1} - 1)^{1/4}}\cr
& \times \Big [ {x\over (2n+1)^{1/2}} + ({x^2\over 2n+1} -
1)^{1/2}\Big ]^{n+1/2}. &(4.7)\cr}
$$

We begin by using the integral representation$^8$
$$
H_n(x) = {2^n\over \sqrt{\pi}} \int^\infty_{-\infty} (x + it)^n
e^{-t^2} dt \eqno (4.8)
$$
twice in (4.1) to obtain \par
$K_N({\bf r}, {\bf r'}) =$ \par
\vskip - 5 mm
$${a({\bf r}) \bar a ({\bf r}')\over \pi
\cosh \mu} {1\over \sqrt{\pi}} \int^\infty_{-\infty} dt_2
e^{-t_2^2}{1\over \sqrt{\pi}} \int^\infty_{-\infty} dt_1
e^{-t^2_1}\sum_{n=0}^{N-1} {\{2(\tanh\mu)(Z + it_1) (\bar Z' +
it_2)\}^n\over n!} \ , \eqno (4.9) $$
where
$$
Z = {z\over \sqrt{\sinh 2\mu}} \qquad . \eqno (4.10)
$$
Following Choquard {\it et al},$^{12}$ we make use of the asymptotic
approximation
$$
\sum_{n=0}^{N-1} {u^n\over n!} \sim {1\over (N-1)!} {u^N\over
u-(N-1)} \Big[1 + O \big({N-1\over (u-(N-1))^2}\big ) \Big ] \eqno
(4.11)
$$
which gives
$$
\eqalignno{K_N({\bf r}, {\bf r'}) & \sim {4a({\bf r}) \bar a ({\bf
r'})\over \pi \cosh \mu} ({1\over 2} \tanh \mu)^N\cr
\times {1\over (N-1)!} {2^{N-1}\over \sqrt{\pi}}
&\int^\infty_{-\infty} dt_2 e^{-t^2_2} (\bar Z' + it_2)^{N-1}
{2^{N-1}\over \sqrt{\pi}}
\int^\infty_{-\infty} dt_1 e^{-t^2_1} ( Z + it_1)^{N-1}\cr
 &\times {(\bar Z' + it_2)( Z + it_1)\over 2 (\tanh \mu) (Z + it_1)
(\bar Z' + it_2) -(N-1)} &(4.12) \cr}
$$
(as in ref. 12, we can show that the terms ignored in going from
(4.9) to (4.12) are $O(1/N)$ relative to the  terms kept).

Consider the double integration in (4.12). Since $N$ is large, by the
method of steepest descent, the slowly varying factor on the last
line can be replaced by its value at that saddle point of each
integration which maximizes the modulus of the integrand on the
second line. Now, the saddle points of the integration over $t_1$
occur when
$$
f'(t) = 0, \qquad \hbox{where} \qquad f(t) = t^2 -(N-1) \ln(Z + it) \ ,
\eqno (4.13)
$$
which gives
$$
t_\pm(Z) = {-Z \pm (Z^2 - 2N + 2)^{1/2}\over 2i}
$$
It can be shown that, for $Z$ describing a point near the boundary
ellipse, taking $t_+ (Z)$ maximizes the modulus of the integrand
(with $(Z^2 - 2N+2)^{1/2}$ defined as the positive square root when
$Z$ is real and larger than $(2N-2)^{1/2}$, and its continuation).
Similarly, $t_+(\bar Z')$ is the saddle point to be chosen for the
integral over $t_2$. Taking the slowly varying factors outside the
integral, with $t_1$ and $t_2$ so specified, we see from (4.8) that
the remaining integrals are precisely the Hermite polynomials, and so
$$
\eqalignno {K_N&({\bf r}, {\bf r'}) \sim {a({\bf r}) \bar a({\bf
r'})\over \pi \cosh \mu} ({1\over 2} \tanh \mu)^N {1\over (N-1)!}\cr &
\times {[\bar Z' + (\bar Z'^2 - 2N + 2)^{1/2}] [Z + (Z^2 - 2 N +
2)^{1/2}] H_{N-1} (\bar Z') H_{N-1}(Z)\over ({1\over 2} \tanh \mu)
[\bar Z' + (\bar Z'^2 - 2 N +2)^{1/2}] [Z + (Z^2 -2 N+2)^{1/2}] -
(N-1)} \ \ . &(4.14)\cr} $$
{}From (4.2), to compute $\rho^{T(b)}_N$, we seek the large-$N$
behaviour of the squared modulus of (4.14), with ${\bf r}$ and ${\bf
r'}$ in the neighbourhood of the boundary  ellipse. Let us
consider the squared modulus of the product of the Hermite
polynomials first. Since, from (4.3) and (4.10),
$$
Z = \sqrt{2N-1} \cosh (\xi + i \eta) ,
$$
(4.7) gives
$$
\eqalignno{
|H_{N-1}&(\bar Z') H_{N-1} (Z)|^2 = {2^{2N-6}[(N-1)!]^4
e^{2(N-1)}\over \pi^2(N-1)^{2N}} {1 \over |\sinh (\xi + i\eta)|}
{1\over |\sinh (\xi' + i\eta')|}\cr
&\times |e^{(N-{1\over 2})[e^{-(\xi+i\eta)} \cosh(\xi + i\eta) + \xi +
i\eta]}|^2 |e^{(N-{1\over 2}) [e^{- (\xi'+i\eta')} \cosh(\xi' + i\eta')
+ \xi' + i\eta']} |^2 \ . & (4.15)\cr}
$$
Furthermore, since
$$
Z + (Z^2 - 2N + 2)^{1/2} = (2N-2)^{1/2} e^{\xi + i \eta} + O({1\over
N})\qquad ,
$$
the squared modulus of the remaining terms in the second line of
(4.14) equals
$$
{4 e^{2(\xi+\xi')} \over |e^{\xi+\xi' + i(\eta-\eta')} \tanh \mu -
1|^2} + O({1\over N}) . \eqno (4.16)
$$
 We also need
$$|a({\bf r}) \bar a({\bf r'})|^2 = \exp\big[-(4N-2) (\sinh
\mu)(e^{-\mu} \cosh^2 \xi \cos^2 \eta + e^\mu \sinh^2 \xi \cos^2
\eta)\Big]$$
$$\times \exp \Big[-(4N-2) (\sinh \mu)(e^{-\mu} \cosh^2 \xi'
\cos^2  \eta' + e^\mu \sinh^2 \xi' \cos^2 \eta')\Big ] .
\eqno(4.17)$$

Substituting (4.17), (4.16), and (4.15) as appropriate in the squared
modulus of (4.14), using Stirling's formula, and substituting the
result in (4.2) gives
$$
\eqalignno{
\rho^{T(b)}_N({\bf r}, {\bf r'}) &\sim - {\tanh \mu \over 8\pi^3(N-1)
\cosh^2 \mu} {1\over |\sinh(\xi + i\eta) \sinh(\xi' + i\eta')|}
{e^{2(\xi + \xi')}\over |e^{\xi+\xi'+i(\eta-\eta')}\tanh \mu -
1|^2}\cr
& \times e^{(N-{1\over 2})[g_\mu(\xi, \eta) + g_\mu(\xi',
\eta')]} \qquad , &(4.18) \cr}
$$
where
$$
\eqalignno{
g_\mu(\xi, \eta) & = \ln \tanh \mu + 1 - 4 (\sinh \mu) (e^{-\mu}
\cosh^2 \xi \cos^2 \eta\cr
& + e^\mu \sinh^2 \xi \sin^2 \eta) + e^{-2\xi} \cos 2\eta + 2\xi
.&(4.19)\cr}
$$

Analogous to the behaviour of $\rho^{T(b)}_N$ for the disk,$^{12}$ we
expect the last exponential in (4.18) to be strongly peaked for ${\bf
r}$ and ${\bf r'}$ in the neighbourhood of the boundary of the
ellipse,
$$
\xi = \xi_b \qquad \hbox{where} \qquad e^{-2\xi_b} = \tanh \mu .
\eqno (4.20)
$$
This is indeed the case as the maximum of (4.19) occurs when $\xi =
\xi_b$. Furthermore, $g_\mu(\xi_b, \eta) = 0$ is independent of
$\eta$. Expansion around this point gives
$$
(N - {1\over 2}) g_\mu (\xi, \eta) \sim -2(\xi - \xi_b)^2 [h(\xi_b,
\eta)]^2 .
$$
Thus, after replacing $\xi, \xi'$ by $\xi_b$  in the
prefactor of (4.18), we obtain
$$
\eqalignno{\rho^{T(b)}_N &\sim -{\tanh \mu\over 8 \pi^3(N-1) \cosh^2
\mu} {e^{4\xi_b} \over |\sinh (\xi_b + i\eta) \sinh (\xi_b + i
\eta')|} {1\over |e^{i\eta} - e^{i\eta'}|^2}\cr
& \times e^{-2(\xi-\xi_b)^2[h(\xi_b, \eta)]^2} e^{-2(\xi' - \xi_b)^2
[h(\xi_b, \eta')]^2} \qquad . &(4.21)\cr}
$$

The task of computing $\rho^{T(b)}_N$ is now completed. To check
(4.6), we substitute (4.21) in (4.5) and note that since the
integrands are sharply peaked at $\xi, \xi' = \xi_b$, we can set
$h(\xi, \eta) = h(\xi', \eta) = h(\xi_b, \eta)$ and remove these
factors from the integrand. Computing the remaining Gaussian
integrations, using (4.4) (where $2N-1$ can be approximated by
$2N-2$) and (4.20) to simplify the resulting  expression, and
recalling that here $e^2/2\pi k_BT = 2$, we see that the predicted
form (4.6) is indeed realized.
\vskip 2cm
\noindent {\bf REFERENCES}
\bigskip
\item {1.} P. Di Francesco, M. Gaudin, C. Itzykson, and F. Lesage,
{\it Int. J. Mod. Phys.} {\bf A9}, 4257 (1994).

\item {2.} A. Alastuey and B. Jancovici, {\it J. Phys. (France)} {\bf
42}, 1 (1981).

\item {3.} J. Ginibre, {\it J. Math. Phys. } {\bf 6}, 440 (1965).

\item {4.} B. Jancovici, {\it Phys. Rev. Lett.} {\bf 46}, 386 (1981).

\item {5.} F. Cornu, B. Jancovici, and L. Blum, {\it J. Stat. Phys.}
{\bf 50}, 1221 (1988).

\item {6.} Ph. Choquard, B. Piller, and R. Rentsch, {\it J. Stat.
Phys.} {\bf 46}, 599 (1987).

\item {7.} C. Cohen-Tannoudji, B. Diu, and F. Lalo\"e, {\it M\'ecanique
Quantique} (Hermann, Paris, 1977).

\item {8.} I.S. Gradshteyn and I.M. Ryzhik, {\it Tables of
Integrals, Series, and Products} (Academic, New York, 1980).

\item {9.} B. Jancovici, {\it J. Stat. Phys.} {\bf 28}, 43 (1982).

\item {10.} B. Jancovici, {\it J. Stat. Phys. } {\bf 80}, 445 (1995).

\item {11.} Ph. Choquard, B. Piller, R. Rentsch, and P. Vieillefosse,
{\it J. Stat. Phys.} {\bf 55}, 1185 (1989).

\item {12.} Ph. Choquard, B. Piller and R. Rentsch, {\it J. Stat.
Phys.} {\bf 46}, 599 (1987).

\item {13.} N. Schwid, {\it Trans. Amer. Math. Soc. } {\bf 37}, 339
(1935).
  \bye